\documentstyle[12pt]{article}
\title{\bf Gravitational conformal invariance and coupling
constants in Kaluza-Klein theory}
\author{F. Darabi$^{a, b, c}$\thanks{e-mail:
f-darabi@cc.sbu.ac.ir} and P. S. Wesson$^{a}$\thanks{e-mail:
wesson@astro.uwaterloo.ca}\\ $^{a}${\small Department of Physics,
University of Waterloo, Waterloo, Ontario, N2L 3G1, Canada.}\\
$^{b}${\small Department of Physics, Tarbiyat Moallem University
of Tabriz, 51745-406, Tabriz, Iran .}\\ $^{c}${\small Institute
for Studies in Theoretical Physics and Mathematics, 19395-1795,
Tehran, Iran. }}
\begin{document}
\maketitle
\begin{abstract}
We introduce a generalized gravitational conformal invariance in
the context of non-compactified 5D Kaluza-Klein theory. It is done
by assuming the 4D metric to be dependent on the extra
non-compactified dimension. It is then shown that the conformal
invariance in 5D is broken by taking an absolute cosmological
scale $R_0$ over which the 4D metric is assumed to be dependent
weakly on the 5th dimension. This is equivalent to Deser's model
for the breakdown of the conformal invariance in 4D by taking a
constant cosmological mass term $\mu^2\sim R_0^{-2}$ in the
theory. We set the scalar field to its background cosmological
value leading to Einstein equation with the gravitational constant
$G_N$ and a small cosmological constant. A dual Einstein equation
is also introduced in which the matter is coupled to the higher
dimensional geometry by the coupling $G_N^{-1}$. Relevant
interpretations of the results are also discussed.
\end{abstract}
\newpage
\section{Introduction}

The theory of conformal invariance has been playing a particularly
important role in the
investigation of gravitational models since Weyl, who introduced
the notion of conformal rescaling of the metric tensor.
Afterwards, it was promoted to the conformal transformations in
scalar-tensor theories, in which another transformation on the scalar
field was required to represent the conformal invariance in modern
gravitational models. There is an open possibility that the gravitational
coupling of matter may have its origin in an invariance breaking effect of
this conformal invariance. In fact, since the ordinary coupling of matter
to gravity is a dimensional coupling (mediated by the gravitational
constant), the local conformal transformations which could change the
strength
of this dimensional coupling, by affecting the local standards of length
and time, are expected to play a key role. In a system which includes
matter,
conformal invariance requires the vanishing of the
trace of the stress tensor in the absence of dimensional
parameters. However,
in the presence of dimensional parameters, the conformal invariance
can be also established for a large class of theories \cite{Bekenstein}
if the dimensional parameters are conformally transformed according to
their dimensions. One general feature of conformally invariant theories
is, therefore, the presence of varying dimensional coupling constants.
In particular, one can say that
the introduction of a constant dimensional parameter into
a conformally-invariant theory breaks
the conformal invariance in the sense that a preferred conformal frame is
singled out, namely that in which the dimensional parameters have
the assumed (constant) configuration.
The determination of the corresponding preferred
conformal frame depends on the nature
of the problem at hand. In a conformally-invariant gravitational model,
the symmetry breaking may be considered
as a cosmological effect. This means that one breaks the conformal
symmetry by defining a preferred conformal frame in terms of the large-scale
characteristics of cosmic matter distributed in a universe with finite
scale factor $R_0$.
In this way, the breakdown of conformal symmetry becomes
a framework in which one can
look for the origin of the gravitational coupling of matter,
both classical \cite{Deser}
and quantum \cite{Salehi}, at large cosmological scales.

The purpose of this paper is to show that one may look for the origins of
both conformal invariance and its breakdown, leading to gravitational
couplings, in a 5-dimensional Kaluza-Klein type gravity theory
\cite{Wesson}. In this popular
non-compactified approach to Kaluza-Klein gravity, the gravitational field
is unified with its source through a new type of 5D manifold in which
space and time are augmented by an extra non-compactified dimension which
induces 4D matter. This theory basically involves writing the Einstein
field equations with matter as a subset of the Kaluza-Klein field
equations without matter \cite{Wesson}, a procedure which is guaranteed by
an old theorem of differential geometry due to Campbell \cite{Cam}.

We show that in the context of pure geometry theory, i.e. $\hat{R}_{AB}=0$
in
5D, one
may
find a generalized conformally-invariant gravitational model.
The well-known conformally-invariant
model of Deser \cite{Deser} in 4D is shown to be a special case when we
drop the dependence of the 4D metric on the extra dimension. Moreover, we show
that the breakdown of conformal
invariance which was introduced in \cite{Deser} by an {\em ad hoc}
non-conformal invariant term inserted into the action naturally
emerges here by
$i)$ assuming a weak (cosmological) dependence of the 4D metric on
the
5th dimension\footnote{This assumption is reasonable since 4D
general relativity is known to be in a very good agreement with present
observations.}
and $ii)$ approximating the scalar field with its cosmological
background value using the well-known cosmological coincidence usually
referred to Mach or Wheeler.

This geometric approach to the subject of conformal invariance and
its breakdown in gravitational models accounts properly for
coupling of the gravitational field with its source in 5D gravity.
It also gives an explanation for the origin of a small
cosmological constant emerging from non-compactified extra
dimension. This subject is the most recent interest in theories
with large extra dimensions \cite{AH}.

The paper is organized as follows: In section 2, we briefly review the
conformal invariant gravitational model and its breakdown in 4 dimensions
due to Deser \cite{Deser}. In section 3, we introduce a generalized
conformal invariant gravitational model in 5 dimensions. In section 4, we
study the breakdown of conformal invariance in 5 dimensions and discuss on
some relevant interpretations. The paper ends with a conclusion.

\section{Breakdown of conformal invariance in 4D}

In this section we briefly revisit the standard work in 4D conformal
invariance due to Deser \cite{Deser}. Consider
the action functional
\begin{equation}
S[\phi]=\frac{1}{2} \int \!d^4 x \sqrt{-g} (g^{\alpha \beta}
\partial_{\alpha}
\phi
\partial_{\beta} \phi +\frac{1}{6} R \phi^2),
\label{1}
\end{equation}
which describes a system consisting of a real scalar field $\phi$
non-minimally coupled
to gravity through the scalar curvature $R$. Variations with
respect
to $\phi$ and $g_{\alpha \beta}$ lead to the equations
\begin{equation}
(\Box -\frac{1}{6} R)\phi=0
\label{2}
\end{equation}
\begin{equation}
G_{\alpha \beta}=6\phi^{-2} \tau_{\alpha \beta}(\phi),
\label{3}
\end{equation}
where $G_{\alpha \beta}=R_{\alpha \beta}-\frac{1}{2}g_{\alpha \beta} R$ is
the Einstein
tensor and
\begin{equation}
\tau_{\alpha \beta}(\phi)= - [\nabla_\alpha \phi \nabla_\beta \phi -
\frac{1}{2}g_{\alpha \beta}
\nabla_\gamma \phi \nabla^\gamma \phi] -
\frac{1}{6}(g_{\alpha \beta}\Box -\nabla_\alpha \nabla_\beta)\phi^2,
\label{4}
\end{equation}
with $\nabla_\alpha$ denoting the covariant derivative. Taking the trace
of (\ref{3}) gives
\begin{equation}
(\Box -\frac{1}{6} R)\phi=0,
\label{5}
\end{equation}
which is consistent with equation (\ref{2}). This is a consequence of the
conformal symmetry of action (\ref{1}) under the conformal transformations
\begin{equation}
\phi \rightarrow \tilde{\phi}=\Omega^{-1}(x) \phi \:\:\:\:\:\:\:\:\:\:
g_{\alpha \beta}\rightarrow \tilde{g}_{\alpha \beta}=\Omega^2 (x)
g_{\alpha
\beta},
\label{6}
\end{equation}
where the conformal factor $\Omega(x)$ is an arbitrary, positive and
smooth
function of space-time.
Adding a matter source $S_{m}$ independent
of $\phi$ to the action (\ref{1}) in the form
\begin{equation}
S = S[\phi] + S_{m},
\label{7}
\end{equation}
yields the field equations
\begin{equation}
(\Box -\frac{1}{6} R)\phi=0
\label{8}
\end{equation}
\begin{equation}
G_{\alpha \beta}=6\phi^{-2}[\tau_{\alpha \beta}(\phi)+T_{\alpha \beta}],
\label{9}
\end{equation}
where $T_{\alpha \beta}$ is the matter energy-momentum tensor.
The following algebraic requirement
\begin{equation}
T=0,
\label{10}
\end{equation}
then emerges as a consequence of comparing the trace of (\ref{9})
with (\ref{8}) which implies that only traceless matter can couple
consistently  to such gravity models.

We may break the conformal symmetry by adding a dimensional mass
term -$\frac{1}{2}\int\!d^4 x \sqrt{-g} \mu^2 \phi^2$, with
$\mu$ being a constant mass parameter, to the action
(\ref{7}). This leads to the field equations
\begin{equation}
(\Box -\frac{1}{6} R+\mu^2)\phi=0
\label{12}
\end{equation}
\begin{equation}
G_{\alpha \beta}+3\mu^2 g_{\alpha \beta}=6\phi^{-2} [\tau_{\alpha
\beta}(\phi)+T_{\alpha
\beta}].
\label{13}
\end{equation}
and we obtain as a result of comparing the trace of (\ref{13}) with
(\ref{12})
\begin{equation}
\mu^2 \phi^2=T.
\label{11}
\end{equation}
Now, the basic input is to consider the invariance breaking as a
cosmological effect. This would mean that one may take $\mu^{-1}$
as the length scale characterizing the typical size of the
universe $R_0$ and $T$ as the average density of the large scale
distribution of matter $ \bar{T}\sim M R_0^{-3}$, where $M$ is the
mass of the universe. This leads, as a consequence of (\ref{11})
to the estimation of the constant background value of $\phi$
\begin{equation}
\bar{\phi}^{-2} \sim R_0^{-2}(M/R_0^3)^{-1} \sim R_0/M \sim G_N,
\label{14}
\end{equation}
where the well-known empirical cosmological relation $G_{N}M/R_0 \sim 1$
(due to Mach or Wheeler) has been used. In order to well-justify the
results we will approximate the correspondence
$\bar{\phi}^{-2} \sim G_N$ with $\bar{\phi}^{-2} \approx
\frac{8\pi}{6}
G_N$. This
estimation for the constant background value of the
scalar field is usually considered in Brans-Dicke type scalar-tensor
gravity
theories. Inserting this background value of $\phi$ into the field
equations (\ref{13}) leads to the following set of Einstein
equations
\begin{equation}
G_{\alpha \beta}+3\mu^2 g_{\alpha \beta}=6\bar{\phi}^{-2}
T_{\alpha \beta} \approx 8\pi G_{N} T_{\alpha \beta}, \label{15}
\end{equation}
with a correct coupling constant $8\pi G_N$, and a term $3\mu^2$
which is interpreted as the cosmological constant $\Lambda$ of the
order of $ R_0^{-2}$. The field equation (\ref{12}) for
$\bar{\phi}$  contains no new information. This is because it is
not an independent equation, namely it is the trace of Einstein
equations (\ref{15}). One may easily check that using $\Box
\bar{\phi}=0$ and $\bar{T}=\mu^2\bar{\phi}^2$, equation (\ref{12})
and the trace of equation (\ref{15}) result in the same equation
as $-\frac{1}{6}R+\mu^2=0$.

\section{5D gravity and generalized conformal invariance}

Consider the 5D metric given by
\begin{equation}
dS^2=\hat{g}_{AB} dx^A dx^B=G\phi^2 g_{\alpha \beta} dx^{\alpha} dx^{\beta}+dl^2
\label{17}
\end{equation}
where the 5D line interval is written as the sum of a 4D part relevant to scalar-tensor
theory and an extra part due to the 5th dimension. The capital Latin indices $A, B, ...$
run over 0, 1, 2, 3, 4, Greek indices $\alpha, \beta, ...$ run over 0, 1, 2, 3, and
five dimensional quantities are denoted by hats. A constant $G$ is also introduced
to leave $G\phi^2$ dimensionless.
We proceed keeping
$g_{\alpha \beta}=g_{\alpha \beta} (x^\alpha, l)$ and $\phi =\phi(x^\alpha)$ as in modern
Kaluza-Klein theory \cite{Wesson}. The metric is general, since we have only used 4 of the
available 5 coordinate degree of freedom to set the electromagnetic potentials,
$g_{4 \alpha}$ to zero.

The corresponding
Christoffel symbols are obtained \begin{equation} \begin{array}{ll}
\hat{\Gamma}^{\alpha}_{\beta \gamma}=\Gamma^{\alpha}_{\beta
\gamma}+\phi^{-1}(\delta^{\alpha}_{\gamma}\nabla_{\beta}\phi+
\delta^{\alpha}_{\beta}\nabla_{\gamma}\phi-g_{\beta
\gamma}\nabla^{\alpha}\phi)
\\
\\
\hat{\Gamma}^{\alpha}_{\beta \alpha}=\Gamma^{\alpha}_{\beta
\alpha}+4\phi^{-1}\nabla_{\beta}\phi
\\
\\
\hat{\Gamma}^{4}_{\beta \gamma}=-\frac{1}{2} \partial_4 \hat{g}_{\beta
\gamma}
\\
\\
\hat{\Gamma}^{\alpha}_{4 \alpha}=\frac{1}{2}\hat{g}^{\alpha \beta}
\partial_4
\hat{g}_{\alpha \beta}
\\
\\
\hat{\Gamma}^{\alpha}_{\beta 4}=\frac{1}{2} \hat{g}^{\alpha \delta}
\partial_4
\hat{g}_{\delta \beta}
\\
\\
\hat{\Gamma}^{4}_{\alpha 4}=\hat{\Gamma}^{\alpha}_{4
4}=\hat{\Gamma}^{4}_{4 4}=0,
\label{18}
\end{array}
\end{equation}
where $\hat{g}_{\alpha \beta}=G\phi^2 g_{\alpha \beta}$. The 5D Ricci
tensor
can
be written in terms of the 4D one plus other terms
$$
\hat{R}_{\alpha \beta}=R_{\alpha \beta}-2\phi^{-1}\nabla_{\alpha}
\nabla_{\beta}\phi+4\phi^{-2}\nabla_{\alpha}\phi
\nabla_{\beta}\phi-\phi^{-2}[\phi \Box\phi+\nabla^{\alpha}\phi
\nabla_{\alpha}\phi]g_{\alpha \beta}
$$
\begin{equation}
\:\:\:\:+\frac{1}{2}G\phi^2[g^{\gamma \delta}
\partial_4 g_{\delta \alpha} \partial_4 g_{\beta
\gamma}-\frac{1}{2}g^{\lambda \delta} \partial_4 g_{\alpha \beta}
\partial_4 g_{\lambda \delta}-\partial_4^2 g_{\alpha \beta}].
\label{19}
\end{equation}
The field equations $\hat{R}_{AB}=0$ then give
\begin{equation}
R_{\alpha \beta}=2\phi^{-1}\nabla_{\alpha}
\nabla_{\beta}\phi-4\phi^{-2}\nabla_{\alpha}\phi
\nabla_{\beta}\phi+\phi^{-2}[\phi \Box\phi+\nabla^{\alpha}\phi
\nabla_{\alpha}\phi]g_{\alpha \beta}-\frac{1}{2}G\phi^2[g^{\gamma \delta}
\dot{g}_{\delta \alpha} \dot{g}_{\beta
\gamma}-\frac{1}{2}g^{\lambda \delta} \dot{g}_{\alpha \beta}
\dot{g}_{\lambda \delta}-\ddot{g}_{\alpha \beta}]
\label{20}
\end{equation}
\begin{equation}
\hat{R}_{4
\alpha}=\nabla_{\alpha}(k^{\alpha}_{\beta}-\delta^{\alpha}_{\beta}
k)=0\:\:\:\:\:with\:\:\:\:\:k^{\alpha}_{\beta}=\frac{1}{2}\hat{g}^{\alpha
\delta} \dot{\hat{g}}_{\delta \beta}=\frac{1}{2}g^{\alpha
\delta} \dot{g}_{\delta \beta}
\label{21}
\end{equation}
\begin{equation}
\hat{R}_{44}=2(\dot{k}-4k^{\alpha}_{\beta} k^{\beta}_{\alpha})=0,
\label{22}
\end{equation}
where an overdot denotes differentiation with respect to 5th coordinate
$l$
( see \cite{W+P} ).
Equation (\ref{20}) may lead to a set of 10 Einstein equations. Equation
(\ref{21}) which have the form of conservation law may also lead to a set
of 4 Gauss-Codazzi equations for the extrinsic curvature
$k^{\alpha}_{\beta}$ of a 4D hypersurface $\Sigma_{l}$ foliating in
5th dimension. Finally, equation (\ref{22}) is one equation for the scalar
combinations of the extrinsic curvature. The Ricci
scalar for the space-time part is obtained by contracting equation
(\ref{20}) with the metric $g_{\alpha \beta}$
\begin{equation}
R=6\phi^{-1} \Box \phi-\frac{1}{2}G\phi^2[g^{\alpha \beta} g^{\gamma
\delta}
\dot{g}_{\delta \alpha} \dot{g}_{\beta
\gamma}-\frac{1}{2}g^{\alpha \beta} g^{\lambda \delta} \dot{g}_{\alpha
\beta}
\dot{g}_{\lambda \delta}-g^{\alpha \beta}\ddot{g}_{\alpha
\beta}].
\label{23}
\end{equation}
Combining equations (\ref{20}) and (\ref{23}) we obtain the Einstein-like
equations with Einstein tensor $G_{\alpha \beta}$ in the left hand side
and some terms of scalar field together with 4D metric and their covariant
derivatives in the right hand side as follows
\begin{equation}
G_{\alpha \beta}=6\phi^{-2} \tau_{\alpha
\beta}(\phi)+\frac{1}{2}G\phi^2[{\cal T}_{\alpha \beta}-\frac{1}{2}{\cal
T}g_{\alpha \beta}]
\label{24}
\end{equation}
where
\begin{equation}
\tau_{\alpha \beta}(\phi)=-\frac{2}{3}\nabla_{\alpha}\phi
\nabla_{\beta}\phi+\frac{1}{6}g_{\alpha \beta} \nabla^{\gamma}\phi
\nabla_{\gamma}\phi-\frac{1}{3}\phi\Box \phi g_{\alpha
\beta}+\frac{1}{3}\phi \nabla_{\alpha}\nabla_{\beta}\phi
\label{25}
\end{equation}
and
\begin{equation}
{\cal T}_{\alpha \beta}=g^{\gamma \delta}
\dot{g}_{\delta \alpha} \dot{g}_{\beta
\gamma}-\frac{1}{2}g^{\lambda \delta} \dot{g}_{\alpha \beta}
\dot{g}_{\lambda \delta}-\ddot{g}_{\alpha \beta}.
\label{26}
\end{equation}
It is easy to show that the tensor $\tau_{\alpha \beta}$ in equation
(\ref{25}) is exactly the same one in equation (\ref{4}).
The field equation for the
scalar field may be obtained by contracting equation (\ref{24}) with
$g_{\alpha \beta}$ or $\hat{g}_{\alpha \beta}$ as
\begin{equation}
\left(\Box -\frac{1}{6}R +\frac{1}{12}G\phi^2 {\cal T}\right)\phi=0.
\label{27}
\end{equation}
We notice that equation (\ref{27}) has a dynamical mass term
$\frac{1}{12}G\phi^2{\cal T}$ with the dimension of $(mass)^{2}$.
In the presence of dimensional parameters, the conformal
invariance can be established for a large class of theories
\cite{Bekenstein} if the dimensional parameters are conformally
transformed according to their dimensions. In this regard,
equation (\ref{27}), although modified by the mass term compared
to (\ref{5}), but is still invariant under the generalized
conformal transformations
\begin{equation}
\phi \rightarrow \tilde{\phi}=\Omega^{-1}(x, l) \phi \:\:\:\:\:\:\:\:\:\:
g_{\alpha \beta}\rightarrow \tilde{g}_{\alpha \beta}=\Omega^2 (x, l)
g_{\alpha
\beta}.
\label{28}
\end{equation}
This is simply because the 5D metric (\ref{17}) is invariant under
the above conformal transformations. Obviously, the following
combination $$ \hat{G}_{\alpha \beta}\equiv \hat{R}_{\alpha
\beta}-\frac{1}{2}\hat{g}^{\gamma \lambda}\hat{R}_{\gamma \lambda}
\hat{g}_{\alpha \beta}\\ \vspace{5mm}=\hat{R}_{\alpha
\beta}-\frac{1}{2}{g}^{\gamma \lambda}\hat{R}_{\gamma \lambda}
{g}_{\alpha \beta} $$ is invariant under (\ref{28}) due to the
invariance of the metric $\hat{g}_{\alpha \beta}$. Therefore,
equation (\ref{24}) which arises as a result of $\hat{G}_{\alpha
\beta}=\hat{R}_{\alpha \beta}-\frac{1}{2}{g}^{\gamma
\lambda}\hat{R}_{\gamma \lambda} {g}_{\alpha \beta}=0$ is
invariant under (\ref{28}). And equation (\ref{27}) as a
consequence of $\hat{g}^{\alpha \beta} \hat{G}_{\alpha \beta}=0$
or ${g}^{\alpha \beta} \hat{G}_{\alpha \beta}=0$ is invariant
under (\ref{28}) as well, regardless of which metric is used to
contraction since the right hand side is zero. Note that although
the initial $l$-independent scalar field $\phi$ transforms to an
$l$-dependent one $\tilde{\phi}$, but the $l$-dependent function
$\Omega^{-1}(x,l)$ will not appear in the transformed scalar field
equation because the metric also transforms in such a way that the
function $\Omega^{-1}(x,l)$ is factored out throughout the
transformed equation rendering the initial scalar field equation.
Therefore, by pure 5D approach we are able to introduce a
generalized conformal invariant gravitational model defined by
equations (\ref{24}), (\ref{27}) and (\ref{28}) subject to the
subsidiary equations (\ref{21}) and (\ref{22}).

\section{Breakdown of conformal invariance in 5D}

Now, we are in a position to compare equations (\ref{27}),
(\ref{24}) with the corresponding equations (\ref{12}),
(\ref{13}). By this comparison it turns out that we are able to
revisit the breakdown of conformal invariance in 4D by a 5D
approach since we have derived the field equations (\ref{27}),
(\ref{24}) which can be identified with (\ref{12}), (\ref{13}) in
the broken phase of the conformal invariance in 4D.

To this end, we take a dimensional analysis. The dimension of
${\cal T}_{\alpha \beta}$ or ${\cal T}$ will no doubt be
$(length)^{-2}$. Now, we assume the cosmological effect
$\dot{g}_{\alpha \beta} \sim \frac{1}{R_0}$ which fixes a very
slow variation of $g_{\alpha \beta}$ over the absolute
cosmological scale $R_0$. This assumption leads to the breakdown
of the conformal invariance since it means that we have fixed our
standard of length by the scale of the universe and that
(comparing equation (\ref{27}) with (\ref{12}) and using
$G\phi^2\sim1$) ${\cal T}$ may be identified with 12$\mu^2$ which
is a constant mass term breaking the conformal invariance. Now, we
put the above identification into the Einstein-like equation
(\ref{24}). We then have
\begin{equation}
G_{\alpha \beta}+3 \mu^2 g_{\alpha \beta}=6\phi^{-2}[ \tau_{\alpha
\beta}(\phi)+\frac{1}{12}\phi^2{\cal T}_{\alpha \beta}],
\label{29}
\end{equation}
which, comparing with equation (13), leads to the identification
\begin{equation}
T_{\alpha \beta}=\frac{1}{12}\phi^2{\cal T}_{\alpha \beta},
\label{30}
\end{equation}
which is the desired result in the context of induced matter theory since
the matter energy-momentum tensor $T_{\alpha \beta}$ is dynamically induced by
the scalar field $\phi$ and higher dimension, namely ${\cal T}_{\alpha
\beta}$. Taking the trace of (\ref{30}) we find \begin{equation}
T=\frac{1}{12}\phi^2{\cal T},
\label{31}
\end{equation}
and by taking ${\cal T}=12\mu^2$ we obtain the equation
(\ref{11}). Now, according to (\ref{31}) we may discuss on the
background value $\bar{\phi}$ corresponding to the absolute
cosmological scale $R_0$. We have already fixed ${\cal T}$ (or
$\mu^2$) by cosmological considerations, namely ${\cal T}\sim
R_0^{-2}$. This was achieved by the 5th coordinate degree of
freedom through $\dot{g}_{\alpha \beta}\sim R_0^{-1}$. The 5th
coordinate degree of freedom accounts for the scalar field in the
general metric (\ref{17}). Thus, (see Eq.(\ref{14}) and the
following discussion) we may take a background value $\bar{\phi}$,
using this coordinate degree of freedom, as
\begin{equation}
\bar{\phi}^{-2} \approx \frac{8\pi}{6} G_N,
\label{32}
\end{equation}
which identifies $G$ with $\frac{8\pi}{6}G_N$ such that
$G\bar{\phi}^2\approx 1$. This condition reduces the general
metric (\ref{17}) to the canonical one \cite{Wesson}. If we now
insert this constant background value $\bar{\phi}^{-2}$ into
equation (\ref{29}) and use (\ref{30}) we find
\begin{equation}
G_{\alpha \beta}+3\mu^2g_{\alpha \beta}\approx8\pi G_N T_{\alpha
\beta}, \label{33}
\end{equation}
in which
\begin{equation}
T_{\alpha \beta}=\frac{1}{16\pi}G^{-1}_N{\cal T}_{\alpha \beta}.
\label{34}
\end{equation}
Equation (\ref{33}) is the well-known Einstein equation in the
broken phase of the conformal invariance with a cosmological
constant $\Lambda=3\mu^2$ and a coupling of matter to gravity,
$G_N$. The scalar field equation (\ref{27}) is the trace of
Einstein equations, so its information is already included in them
(see the discussion in section 2).

Now, the relevance of 5D approach manifests. It relates the
current upper bound value of the cosmological constant $\Lambda
\sim R_0^{-2}$ to a geometric phenomenon in which {\em the
cosmological constant is generated by the very slow variation of
4D metric with respect to 5th dimension}\footnote{In a recent work
of Arkani-Hamed {\em et al} \cite{AH}, a small effective
cosmological constant is emerged from a large extra dimension in a
non-compactified approach to 5D Kaluza-Klein gravity. Also, in a
compactified model of Kaluza-Klein cosmology \cite{DS}, smallness
of the cosmological constant is related to smallness of the
compactified dimension. Therefore, it seems that the subject of
cosmological constant in higher dimensional (at least in 5D)
models is inevitably involved with extra dimension.}. Moreover, it
unifies the origins of the matter and the cosmological constant in
that they appear as ``two manifestations of higher dimensional
geometry''.

The traditional Einstein equation (\ref{33}) may alternatively be
written in its pure geometric form
\begin{equation}
G_{\alpha \beta}+3\mu^2g_{\alpha \beta}\approx \frac{1}{2}{\cal T}_{\alpha
\beta},
\label{35}
\end{equation}
in which the coupling constant $G_N$ is removed from theory. To
say, although the Einstein tensor $G_{\alpha \beta}$ couples to
the matter $T_{\alpha \beta}$ by $G_N$ but the matter itself
couples by $G_N^{-1}$ to the geometry ${\cal T}_{\alpha \beta} $
(\ref{34}) and so the coupling $G_N$ is removed. In this level,
the appearance of $G_N$ in the traditional Einstein equation seems
to be a mathematical tool only for dimensional consistency.
However, in the physical level equations (\ref{33}) and (\ref{34})
exhibit an interesting phenomenon, with varying $G_N$, in that if
$G_N$ decreases with time leading to a weakly coupling of gravity
$G_{\alpha \beta}$ to the matter $T_{\alpha \beta}$ (\ref{33}),
the matter itself will then be coupled strongly to the hidden
geometry ${\cal T}_{\alpha \beta}$ (\ref{34}). Regarding the
present small value of $G_N$ we find an strong coupling of matter
$T_{\alpha \beta} $ to the higher dimensional geometry ${\cal
T}_{\alpha \beta}$. This strong coupling may account for
non-observablity of the 5th dimension. In other words, the effects
of the 5th dimension may be hidden behind this strong coupling and
what we observe as the matter may be {\em the manifestation of a
weak effect of 5th dimension which is strengthened by a strong
coupling} $G_N^{-1}$. This means that going back in time in $G_N$
varying scenarios we will encounter with an era $G_N \sim 1$ in
which $T_{\alpha \beta} $ may decouple from ${\cal T}_{\alpha
\beta}$ leading to a naked geometry of 5th dimension without the
concept of matter, as indicated in Eq.(\ref{35}). In conclusion it
may be said that two equations (\ref{33}) and (\ref{34}) define
{\em dual weak-strong} regimes, in 5D approach to coupling
constants, and that equation (\ref{34}) defines a {\em
dual}-Einstein equation coupling matter to higher dimension.

It is worth noting that the conformal invariance in 4D may be easily
recovered
in this 5D approach by restricting the 4D metric $g_{\alpha \beta}$
to be independent of 5th dimension (simply by assuming Kaluza-Klein
compactification condition for
higher
dimension). The relevant field equations in this choice
are
\begin{equation}
R_{\alpha \beta}=2\phi^{-1}\nabla_{\alpha}
\nabla_{\beta}\phi-4\phi^{-2}\nabla_{\alpha}\phi
\nabla_{\beta}\phi+\phi^{-2}[\phi \Box \phi +\nabla^{\alpha}\phi
\nabla_{\alpha}\phi]g_{\alpha \beta},
\label{36}
\end{equation}
where by taking the trace of (\ref{36}) and combining it with (\ref{36})
we obtain the conformal invariant equations (\ref{2}) and (\ref{3}).
The origin of this conformal invariance in 4D turns out to be the
invariance of the 4D part of 5D metric
\begin{equation}
\hat{g}_{AB}(x^A)= \left(\begin{array}{cc} G\phi^2(x^{\alpha})g_{\alpha
\beta}(x^{\alpha}) & 0 \\ \\ 0 & 1
\end{array}\right)_,
\label{37}
\end{equation}
under the conformal transformations (\ref{6}).

\section*{Conclusion}

A key feature of any fundamental theory consistent with a given
symmetry is that its breakdown would lead to effects which can
have various manifestations of physical importance. Therefore, in
the case of conformal symmetry in gravitational models, one would
expect that the corresponding cosmological invariance breaking
would have important effects generating the gravitational coupling
and the cosmological constant. In this paper we have introduced a
generalized conformally-invariant gravitational model of 5D
gravity theory $\hat{R}_{AB}=0$, with 4D part that is dependent on
the extra dimension. The conformal invariance in 4D then becomes a
special case when we take the 4D metric to be independent of the
extra dimension. Moreover, we have shown that the cosmological
breakdown of conformal symmetry in a conformally-invariant
gravitational model in 4D may be naturally derived in this context
if we assume a weak (cosmological) dependence of the 4D metric on
the higher dimension and use the cosmological coincidence due to
Mach or Wheeler to approximate the scalar field by its
cosmological background value. This approach to the issue of
couplings and parameters in gravity leads to a geometric
interpretation for the small cosmological constant $\Lambda$.
Moreover, a dual coupling $G_N^{-1}$ is introduced by which the
matter couples strongly to the geometric effects of higher
dimension through a {\em dual} Einstein equation, and
non-observability of higher dimension is then justified.

We also mention to the generality of the 5D conformal invariance.
In Deser's model the conformal symmetry is broken once a {\em
constant mass term} is introduced. However, in 5D approach a {\em
dynamical mass term} is appeared without breaking the conformal
symmetry. This generalized symmetry is broken when we take a
preferred conformal frame by introducing an {\em absolute length
scale} $R_0$ through $\dot{g}_{\alpha \beta} \sim R_0^{-1}$. In
other words, what we call the conformal invariance in Deser's
model is not really a conformal invariance; it is just {\em scale
invariance} which is a special case of conformal invariance. This
is because the dimensional constant mass term could not transform
conformally\footnote{The conformal invariance is more general than
scale invariance which is used in Deser's model. If scale
invariance is characterized by vanishing of the trace of the
energy-momentum tensor, conformal invariance implies scale
invariance in the absence of dimensional parameters in the
theory.}.

There is a natural question in the context of induced matter
theory about its possible connection to quantum theory. This is
because we can induce the matter geometrically from the 5th
dimension whereas we know the matter has a underlying quantum
structure. Therefore, it deserves to pay attention to this issue.
First, it is well-known that the existence of a dimensional
gravitational constant $G_N$ is the main source of
non-renormalizablity of quantum gravity. On the other hand, the
quantum theory approach to the traditional Einstein equation
suffers from the problem that the left hand side is geometry and
the right hand side is the matter. Equation (\ref{35}), however,
as a pure geometric Einstein equation is free of $G_N$. Moreover,
both sides of this equation has the geometric structure. Perhaps,
it is helpful to study the 4D quantum gravity in this pure
geometric 5D approach. Second, in a study of 4D conformal
invariance in QFT in \cite{Salehi}, the following equation like
our scalar field equation (\ref{27}) is obtained $$
\left(\Box-\frac{1}{6}+\phi^{-2} {\cal S}^{\alpha}_{\alpha}
\{\omega\} \right)\phi=0 $$ in which ${\cal
S}^{\alpha}_{\alpha}\{\omega\}$ is the trace of the tensor ${\cal
S}_{\alpha \beta}\{\omega\}$ describing the distribution of matter
due to local quantum effects. It is therefore very appealing to
think about how the higher dimensional effects in 5D may play the
role of quantum effects in 4D.

\section*{Acknowledgment}

F. Darabi would like to thanks B. Mashhoon, W. N. Sajko and L. de Menezes
for useful comments.

\end{document}